\documentclass[12pt,preprint]{emulateapj}
\usepackage[latin9]{inputenc}
\usepackage{amsthm}
\usepackage{amsmath}
\usepackage{graphicx}
\usepackage{color}

\begin{document}

\title{Effects of Intermediate Mass Black Holes on Nuclear Star Clusters}

\author{Alessandra Mastrobuono-Battisti$^1$, Hagai B. Perets$^1$ \&
Abraham Loeb$^2$} 

\affil{$^1$ Physics Department, Technion-Israel Institute of Technology, Haifa,
Israel 32000; $^2$ Institute
for Theory and Computation, Harvard University, 60 Garden Street,
Cambridge, MA 02138, USA}

\begin{abstract}
Nuclear star clusters (NSCs) are dense stellar clusters observed in
galactic nuclei, typically hosting a central massive black hole. Here we study the possible formation 
and evolution of NSCs through the inspiral of multiple star clusters hosting intermediate mass black
holes (IMBHs). Using an N-body code we examine the dynamics of the
IMBHs and their effects on the NSC. We find
that IMBHs inspiral to the core of the newly formed NSC and segregate
there.  Although the IMBHs scatter each other and the stars, none of them is ejected from the
NSC. The IMBHs are excited to high eccentricities and their radial
density profile develops a steep power-law cusp. The stars also develop a power-law cusp  (instead of the
central core that forms in their absence), but with a shallower slope. 
The relaxation rate of the
NSC is accelerated due to the presence of IMBHs, which act as massive-perturbers.
This in turn fills the loss-cone and boosts the tidal disruption rate
of stars both by the MBH and the IMBHs to a value excluded by rate estimates based on current observations. 
{  Rate estimates of tidal disruptions can therefore provide a cumulative constraint on the existence of IMBHs in NSCs.}
\end{abstract}
\maketitle

\section{Introduction}

Nuclear stellar clusters (NSCs) are dense and compact stellar systems
observed in galactic nuclei, many of which hosting a massive black
hole (MBH) at their center \citep{Bo10}. Less massive galaxies seem to
contain an NSC but not always an MBH, while the most massive ones show
the presence of a central MBH \citep[see e.g.][]{Gr08, GS09, Ji11, SG13}. 
Our own Galactic center (GC) hosts both an NSC and a central MBH, Sgr
A$^{*}$ of mass ${\rm M_{BH}\sim4\times10^{6}}$ ${\rm M_{\odot}}$.

Two main hypotheses have been suggested to explain the origin of NSCs:
\textit{(i)} the in-situ formation, where molecular gas coming from
more external regions is channeled to the GC leading to
an episodic star formation epoch \citep{Mi04}; and \textit{(ii)} the
cluster-infall/merger scenario, where dense massive stellar clusters
decay toward the center of their host galaxy due to dynamical friction,
merge and form the NSC. These two scenarios are not mutually
exclusive, and both can contribute to the formation of NSCs. Here we
focus only the cluster-infall scenario.

The merger scenario was proposed by \citet{Tr75}, and later explored
in many works in the case of a generic galaxy without a central MBH
\citep[see e.g.][]{Ca93,Ag11,Ca08a,Ca08b}. Other studies explored the
case of galaxies hosting an MBH \citep{An13}, and, in particular, the
origin of the NSC in the GC of the Milky Way
\citep{An12,Gn13,PM14}. Relaxed stellar systems around an MBH give rise to 
a  cusp structure   (a power-law distribution with a slope of
$-7/4$, for a single mass population and somewhat shallower slope for
most stars in a multi-mass population, \citealt{Ba76}).  However, the relaxation time
of the system may be longer than the age of the stars in the NSC, in
which case such a cusp may not form. The two-body relaxation time of
the observed Galactic NSC, evaluated at the influence radius of the
MBH ($2$-$3$~pc) is estimated to be $\sim20$-$30$~Gyr \citep{Me10}.
Current observations of the GC suggest the existence of a flat core in
the distribution of the old stellar population of the NSC.

The observed features of the Milky Way NSC have been well reproduced
by \citet{An12} in their N-body simulations where they studied 12
consecutive inspirals of massive, compact clusters, which build-up an
NSC with a central core of $\sim0.5$~pc and and external $\sim
r^{-1.8}$ power law outside. Recent studies suggest that this scenario
may also lead to age/color and mass segregation in the NSC that can
potentially be observed \citep{PM14}.  In analogy with galaxies hosting a central
MBH, \citet{Si75} proposed that massive clusters may host an
intermediate mass black hole (IMBH) at their center. The formation of
IMBHs in dense clusters has been explored in several studies
\citep{Lo94,Ma01, Eb01,Br03}.  Numerical simulations of the formation and
evolution of young dense clusters show that runaway collisions between
stars can produce a massive star ($\sim10^{2}-10^{4}$~M$_{\odot}$)
that sinks to the center of the cluster due to dynamical friction
\citep{GFR04,Po04, FGR06,Po06,Fu09}, and later collapses to form an
IMBH with a mass of $\sim10^{2}-10^{4}$~M$_{\odot}$.  Given the possible
existence of such IMBHs in massive clusters, the cluster-infall
scenario may be significantly affected by such objects.  Here we extend
the cluster-infall scenario to include this possibility, and study the
effects of IMBHs on the NSC formation and evolution. 

We use N-body simulations similar to those described in \citet{An12},
but populate the infalling clusters (ICs) with IMBHs. The outline of the
paper is as follows. We begin with a description of the initial
conditions and our methods in \S \ref{sec:ICs}. We then present
our results (\S \ref{sec:res}), discuss their implications
(\S \ref{sec:disc}) and summarize (\S \ref{sec:sum}).

\section{Initial Conditions and Methods}

\label{sec:ICs} 

In our simulations we used the same models and methods described in
Paper I \citep{An12}, where a detailed description of the initial
conditions of initial ICs, as well as the galaxy model for
the background stellar population can be found. In brief, we followed
the decay and the merging of 12 massive dense clusters inside the
Galactic bulge by means of fully self-consistent N-body
simulations. The bulge of the simulated galaxy has been modeled based
on recent observations \citep{La02}. A central MBH as massive as the
SgrA* is placed at the center of the Galaxy.
{ A truncated power-law model has been used for this component:
\begin{equation}\label{dm}
\rho_{g}(r)=\tilde{\rho}\left(\frac{r}{\tilde{r}}\right)^{-\gamma}\text{sech}\left(\frac{r}{r_{\rm cut}}\right).
\end{equation}
where $\tilde{\rho}=400$M$_{\odot}/$pc$^{3}$ and $r_{\rm cut}=22~$pc, thus the galactic mass is $9.1\times10^7\text{M}_\odot$.
{ Finally, as in Paper I,  $\gamma$ is set to $1/2$}.

The Monte Carlo method described in \citet{SM05} has been used to sample the 
particles, {whose mass is fixed to be}  $400~M_\odot$, as to have $N_g=227523$ super-stars.}

The IC modeling follows Paper I with the addition of a central
IMBH of $10^{4}M_{\odot}$.  
In particular, we used a highly compact and initially massive tidally truncated King model (see Paper I for details on the truncation method). The central velocity dispersion of the model is $\sigma_0=35$~km s$^{-1}$, the
core radius $r_c=0.5$~pc, and $W_{0}=8$. After the truncation the cluster mass is $\sim 1.1\times 10^6~M_\odot$. We sampled the model using $N_{IC}=5715$, 200~$M_\odot$ particles.

The ICs are compact and their cumulative
mass ($\sim1.5\times10^{7}M_{\odot}$) is comparable to the mass of the
Galactic NSC \citep{Ge10}. Massive ICs decline to the galactic nucleus
in a time significantly shorter than a Hubble time \citep{An12,Ca93}.
The clusters are given time to relax in isolation before they are put
into the cluster-infall simulation. The relaxation timescale is
relatively short and is meant to assure the stability of the system.

\begin{figure}
\centering
\includegraphics[width=0.4\textwidth]{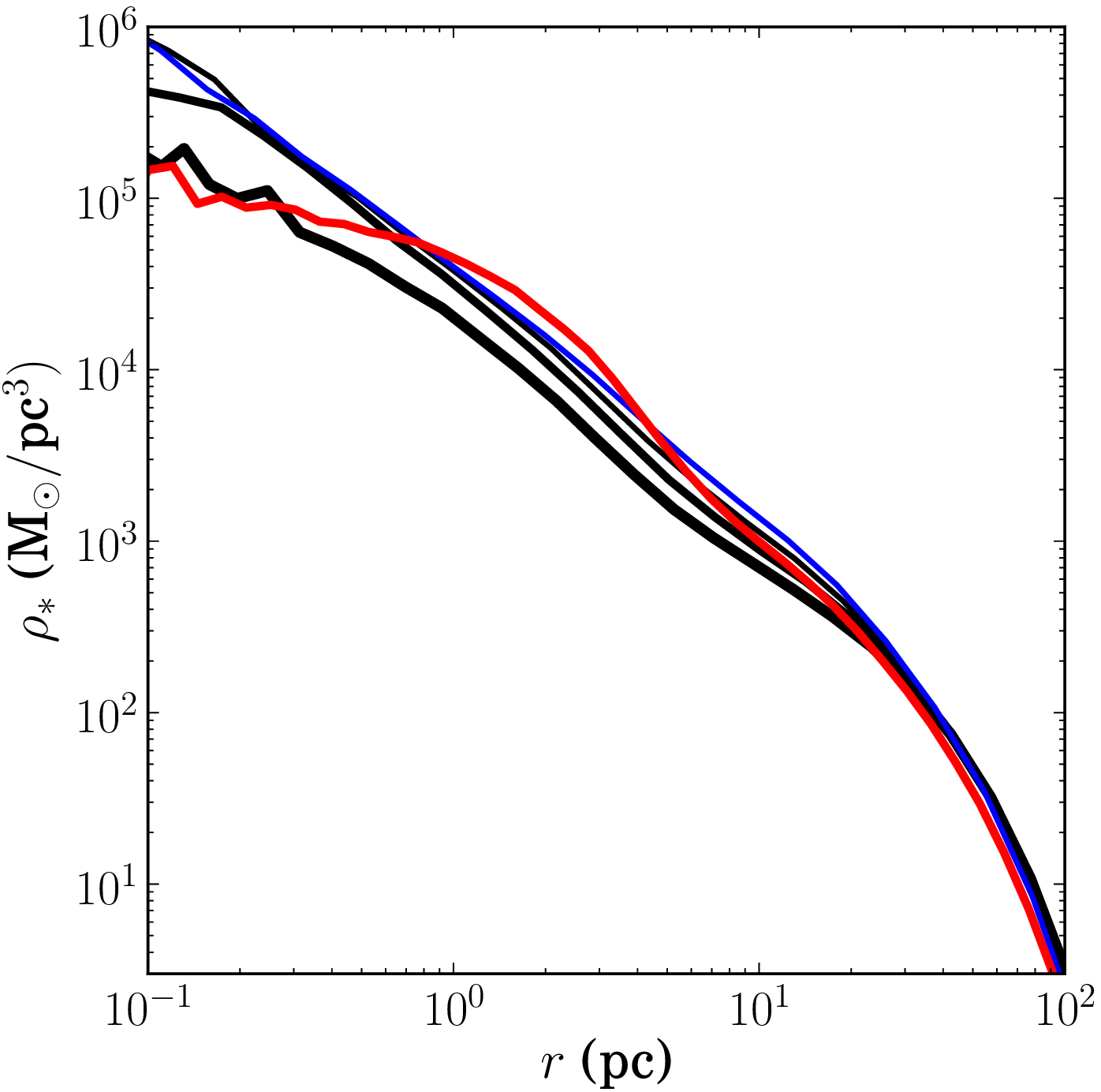}
\includegraphics[width=0.4\textwidth]{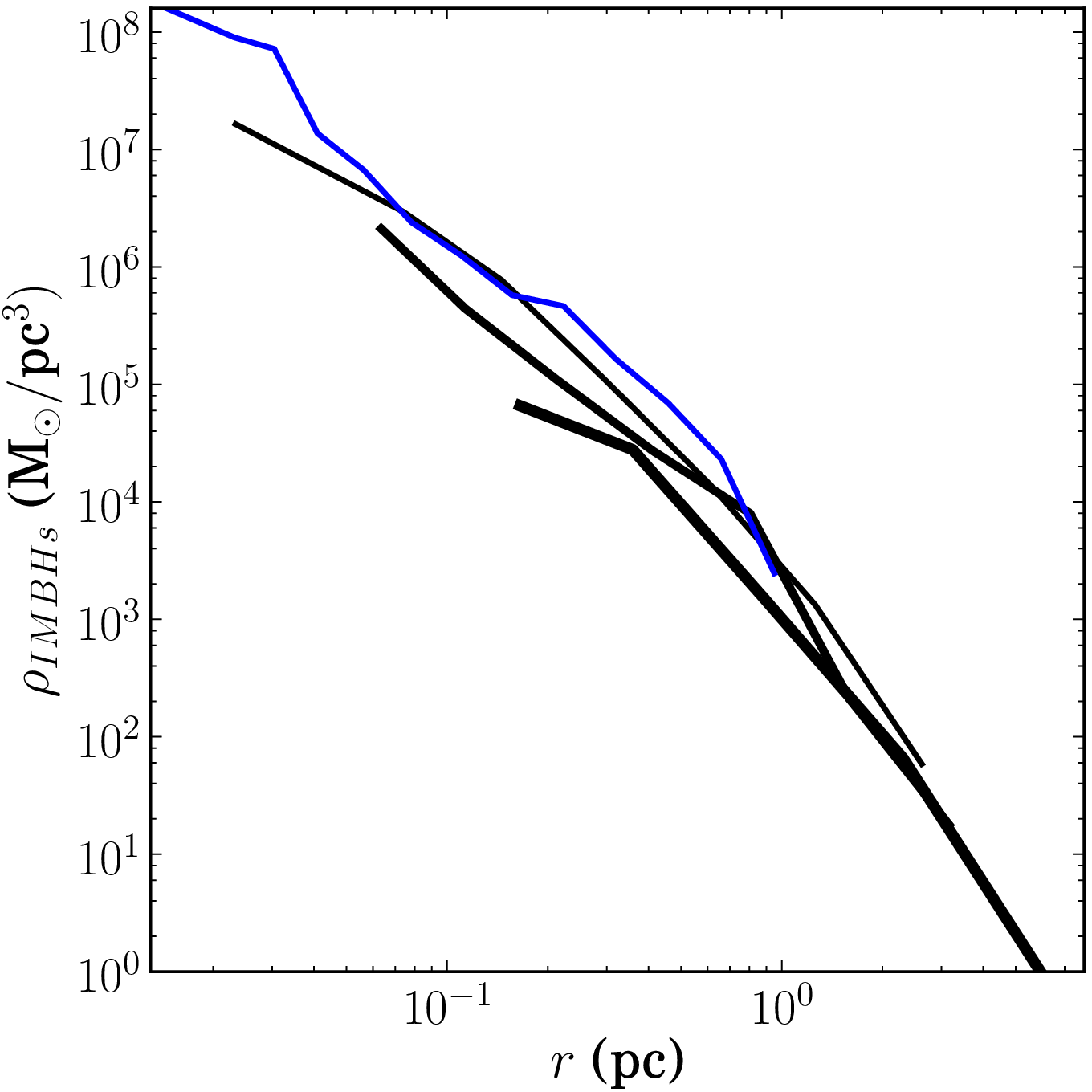}
\caption{{ Left panel}: The spatial density profile of the stellar
component of the growing NSC after 3, 6, 9 and 12 decays. The
thickness of lines decreases with time { and the blue solid line is
the density profile after 12 merging events}. The red solid line represents
the density profile of the NSC obtained after 12 merging events in
Paper I. With IMBHs the NSC develops a density cusp whereas without
IMBHs the NSC has a central core. { Right panel}: same for the
density profile of IMBHs.}
\label{fig:den_ev} 
\end{figure}

We let 12 ICs inspiral toward the GC in a consecutive
order. A new IC was sent only after the previous one has merged at the
center of the Galaxy and has reached a quasi-stationary state. We
follow the ICs after they have already reached the central region of
the galaxy. Each IC is initially placed on a circular orbit with a
radius of $20$~pc from the MBH, as described in \citet{Gu09} and in
Paper I.\\
\begin{figure}
\centering
\includegraphics[width=0.45\textwidth]{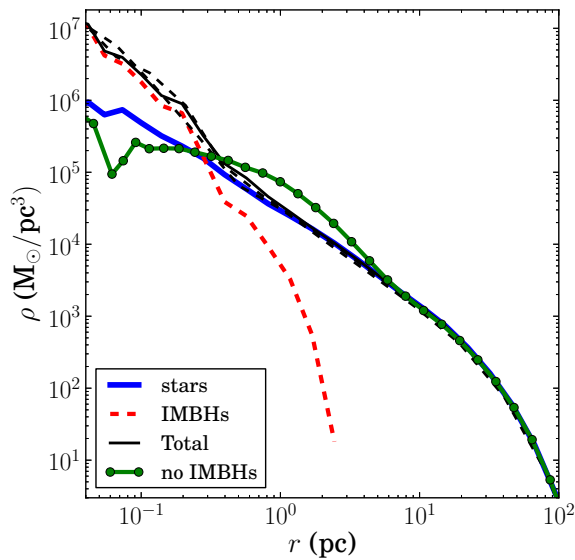}
\caption{Density profiles of the NSC at the end of the simulation
(black solid line). { We run two additional simulations 
to check our results. The density profiles of the NSCs obtained in these cases are shown by the black {dashed} lines.} The blue solid line represents the spatial profile of
the stellar component, while the red dashed line is the final density
profile of the 12 IMBHs. The green solid line with bullets shows the
density profile of the NSC without IMBHs (from Paper I). The IMBHs
introduce a central cusp due to enhanced relaxation.}
\label{fig:denIMBHs} 
\end{figure}

We ran our simulations with the $\phi GRAPE$ code \citep{Ha07}, a
direct-summation code optimized for computer clusters accelerated by
GRAPE boards \citep{Ma98}. The time integrator for this code is a
fourth-order Hermite with a predictor-corrector scheme and
hierarchical time stepping. $\phi GRAPE$ has been recently adapted to
run on graphic processing units (GPUs) by means of Sapporo, a CUDA
library that emulates double-precision force calculations on
single-precision hardwares \citep{Ga09}.  The accuracy and performance
of the code are set by the time step parameter $\eta$ { (the time-step is
proportional to the square root of $\eta$)} and the
smoothing length $\epsilon$. In our simulations we chose $\eta=0.01$
and $\epsilon=0.01$~pc$=2\times10^{-2}r_{c}$. With this choice we have
a relative energy conservation of less than $10^{-4}$ { until the tenth merging event, and $\sim 10^{-3}$ for the remaining part of the simulation. This decay in energy conservation is probably due to the interaction between stars and IMBHs and between the IMBHs}. The simulations have been run on the GPU partition of the
Tamnun cluster  at the Technion. This
partition consists of four nodes, each equipped with one NVidia Tesla M2090. Since $\phi GRAPE$ can run on multiple GPUs only if they are installed on the same node, we used one GPU for each run. Each simulation needed at least four months to be completed.  
Due to computational limitations the ICs were represented by a smaller number of particles than realistic systems, each with a higher particle mass as to represent the total cluster mass (as discussed above). 
 The simulation time was then rescaled to real time using the procedure described in detail by \cite{MBP13} and \cite{PM14}. 

\begin{figure}
\centering
\includegraphics[width=0.5\textwidth ]{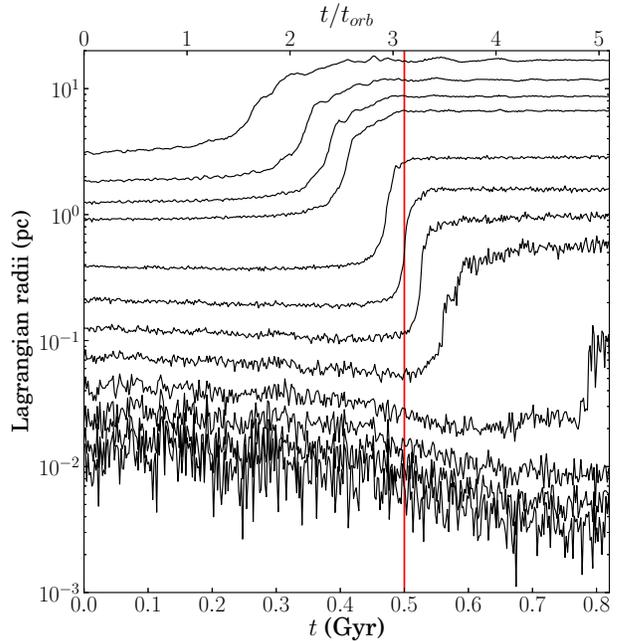} 
\caption{{ Lagrangian radii of stars from the first decaying cluster in respect to the center of density of the cluster itself. {The percentage of mass enclosed within each of the plotted radii
are: 0.025, 0.06, 0.16, 0.4, 1, 2.5, 6.3, 16, 40, 50, 63 and 79\%}. The time given on the bottom x-axis is rescaled with the mass of the particles and softening length used. On the top x-axis we show the time in terms of the period of the circular orbit at 20 pc, i.e. the starting distance of the ICs. The red vertical line shows when the center of density of the system is at 2pc from the central MBH. This time roughly corresponds to the disruption time of the cluster.}}\label{fig:lag}
\end{figure}

We followed the inspiral of the 12 ICs toward the GC
until they were tidally destroyed by the MBH and showed a quasi-stationary state. 
{ In order to support the reliability and reproducibility of our results {we have run two additional simulations using, as initial conditions, 
different samplings of the same initial physical conditions}. 
Here we show only results from the first run, but these were verified with the additional runs, both showing very similar results. In Figure \ref{fig:denIMBHs} the density profile for the final NSC is shown for all three simulations. As it is apparent from the plot, the curves do not show any significant difference supporting our conclusions.}
{ Figure \ref{fig:lag} shows the Lagrangian radii of the first decaying cluster, evaluated in respect to the center of density \citep{CH85} of the cluster itself.  After the cluster is destroyed, we waited for the Lagrangian radii to be almost constant before
sending the following cluster. }
After the end of all the infalls we followed the
relaxation process of the final system. This relaxation is driven by
two-body interactions between stars, stars and IMBHs and among the
IMBHs.

\section{Results}

\label{sec:res}

\subsection{The Density Profile and Kinematic Properties}

To follow the formation of the central NSC, we plot the density
profile after each merging. Figure \ref{fig:den_ev} shows the density
profile of the stellar component of the system (galaxy+stars in the
clusters, left panel) and of the IMBHs (right panel) after 3, 6, 9 and
12 merging events. 
While the background bulge does not change its
profile, the central region evolves toward an NSC
structure. The presence of the IMBHs significantly changes the
properties and evolution of the NSC.  In particular, in the
simulations run without the central IMBH \citep{An12} the ICs are
disrupted at $\sim2$~pc {( see the red line in Figure \ref{fig:lag})}, and their stars develop a central flat core
(see the red solid line in the left panel of Figure
\ref{fig:den_ev}). In comparison, following the disruption of the
IMBH-hosting ICs, the IMBHs keep inspiraling to the center, and lead
to the fast relaxation of the NSC stellar population, giving rise to a
central cusp, with a somewhat shallower slope than a \citet{Ba76}
profile, likely due to the strong mass segregation induced by the
IMBHs (see \S \ref{subs:sms}). As can be seen in the left panel
of Figure \ref{fig:den_ev}, a nearly flat core with radius $\sim1$~pc still
exists, but by the sixth IC infall a central cusp has already formed,
both in the stellar and IMBHs components (right panel of Figure
\ref{fig:den_ev}).

At the end of the last infall 
{ the inner slope of the radial density profile for the whole system (Galaxy+NSC) is $\sim -2.1$,} i.e., significantly different than the density profile of the IMBH-free
cluster infall scenario \citep{An12, PM14}, where the NSC shows a
central core. Although the core size decreases with each infall, it
still exists even after the long-term evolution and further relaxation
of the system. Indeed, the IMBHs act as massive perturbers, and
thereby significantly decrease the relaxation time of the host system
\citep{Pe07}.  At the end of the simulation, i.e. after 12Gyr {($\sim 400$ internal crossing times of the ICs)}, both the stellar cusp and steep radial distribution of the IMBHs are fully
developed (see Figure \ref{fig:denIMBHs} and \S \ref{subs:sms} for more
details).

The kinematic properties of the NSC in the cluster-infall with the IMBHs
simulations are similar to those found in Paper I. At the end of the
last merger, the NSC is tangentially anisotropic between $0.3$ and $30$~pc,
and it is almost isotropic outside this radius. Since the cluster
evolves toward isotropy, the tangential anisotropy decreases in the
subsequent 2 Gyr of evolution. This is also seen in the axial ratios
and in the triaxiality parameter: the cluster remains oblate even after
the mergers evolution, within $30$~pc. As found earlier, outside this
radius the system retains its initial spherical configuration.

\begin{figure}
\centering
\includegraphics[width=0.5\textwidth]{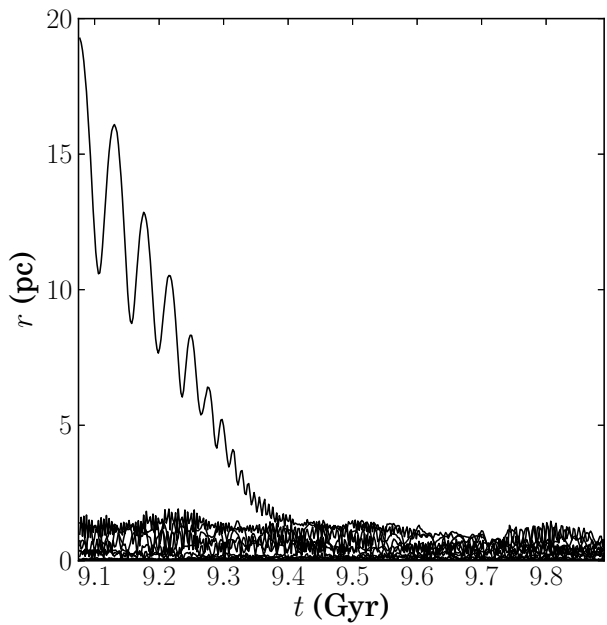}
\caption{Upper line shows the temporal decay of the orbital radius
of the last IMBH to fall in. The radius vs. time of the other 11
IMBHs is also shown.}
\label{fig:last} 
\end{figure}

\subsection{Dynamical Evolution of the IMBHs}

As shown in the previous section, the presence of IMBHs significantly
affects the inner structure of the NSC. Figure \ref{fig:last} shows
the galactocentric distance of the last IMBH to decrease as a function
of time. The IMBH initially decays with its host IC, until the cluster
is tidally disrupted. The IMBH then continues to inspiral by itself
due to dynamical friction induced by the NSC stars. The other curves
in Figure \ref{fig:last} represent the distance from the GC
of the previous IMBHs that already settled to the
center. Though the IMBHs interact and scatter each other, they are
never ejected from the system and are contained within the central 2 pc
during the post-merger evolution.

\begin{figure}
\centering

\includegraphics[width=0.5\textwidth]{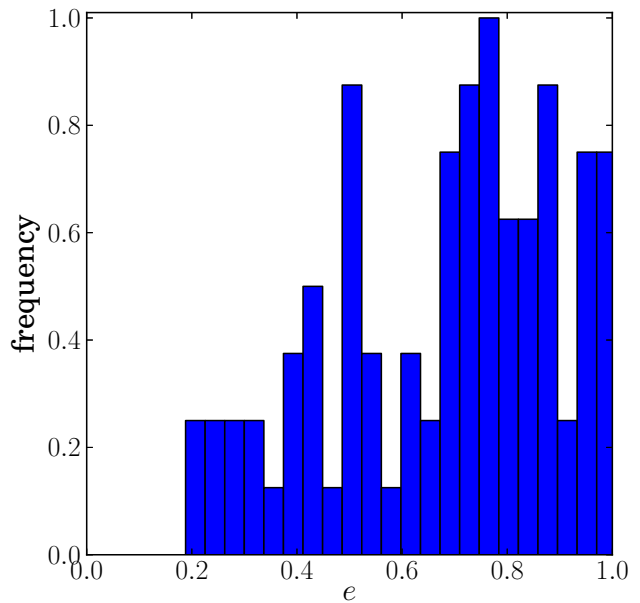}
\caption{Eccentricity distribution of the IMBHs. Multiple snapshots
close in time have been used to increase the data resolution. }
\label{fig:ecc} 
\end{figure}

Figure \ref{fig:ecc} shows the eccentricity distribution of the
IMBHs at the end of the simulation. The IMBHs can reach high
eccentricities, with values close to unity due to mutual scatterings.

\section{Discussion}

\label{sec:disc}

\subsection{Fast Relaxation and High Rate of Tidal Disruption Events}

The presence of IMBHs in ICs significantly affects the formation and
evolution of NSCs. The inspiral of the IMBHs to the nucleus of the NSC both
provides additional stars to the central region of the NSC, and leads
to its faster relaxation. Acceleration of relaxation processes can
enhance the rate of scattering stars into highly eccentric orbits and the 
refilling of the loss cone for tidal disruption events (TDEs) of
stars \citep[see also][]{ML12}. Following \citet{Pe07} we evaluated the local empty and full
loss cone rates, { using stellar number density profiles based on the mass density profiles from our simulations (assuming the stellar mass divided among Solar mass stars)}
\begin{equation}
\frac{d\Gamma_{e}}{d\log r}\sim\frac{N_{\star}(<r)}{\ln(J_{c}/J_{lc})t_{r}(r)},
\end{equation}

\begin{equation}
\frac{d\Gamma_{f}}{d\log r}\sim\frac{J_{lc}^{2}}{J_{c}^{2}}\frac{N_{\star}(<r)}{P(r)},
\end{equation}
where $N_{\star}(<r)$ is the number of stars enclosed within $r$,
$t_{r}(r)$ is the relaxation time at $r$, $J_{lc}$ and $J_{c}$
are, respectively, the loss cone angular momentum and the angular momentum
of the circular orbit of radius $r$ and, finally, $P(r)$ is the
Keplerian period at $r$.

The total loss cone refilling rate is then
\begin{equation}
\Gamma_{tot}=\int\frac{d\Gamma}{dr}dr=\int\min\left[\frac{d\Gamma_{f}}{dr},\frac{d\Gamma_{e}}{dr}\right]dr.
\end{equation}

The presence of IMBHs in the NSC decreases the relaxation time by a
factor $\propto N_{I}M_{I}^{2}/(N_{\star}M_{\star}^{2})$, where
$N_{I}$ and $M_{I}$ are the number and the mass of the IMBHs,
while $N_{\star}$ and $M_{\star}$ are the corresponding number and
mass of the stars. The relaxation time is, therefore, shortened by a few
orders of magnitude, allowing for the rapid relaxation of the
system. This also significantly affects the loss cone refilling rate
for the tidal disruption of stars by the MBH. With the IMBHs, the
system is in the full loss cone regime throughout all of its radial
extent, while, without them, the loss-cone is empty inside the radius of
influence of the MBH. The integrated TDE rate
with the IMBHs is $10^{-3}\:{\rm yr^{-1}}$, while otherwise,
accounting only for stars, we find a rate of $1.6\times10^{-5}$ ${\rm
yr}^{-1}$, comparable to the typical rate estimated for NSCs similar
to that of the Milky Way \citep{Me13}.  The inspiral of IMBHs can,
therefore, increase the TDE rate by two orders of magnitude.
{ It is interesting to notice that a similar result has been
obtained by \cite{XC13} using purely analytical calculations.}

The inferred rate of TDEs from observations is of the order of
$10^{-5}-10^{-4}\,{\rm yr^{-1}}$ \citep[see][and references
therein]{Ko12, KS14}, much lower than expected for multiple-IMBHs hosting
NSCs such as those seen in our simulations. We, therefore, conclude that
typical NSCs should not contain IMBHs. If cluster infall is one of the
main channels for NSC formation, this also implies that the majority of
massive clusters  may not host IMBHs.
{ We do caution that the observational rate estimates are still inaccurate; better data will be available from future large surveys.}

The presence of an IMBH could also induce an enhancement in the number of stars plunging to the MBH through secular Kozai evolution, if the IMBH reaches close enough to the MBH as to provide a significant perturbation, not quenched by the mass precession from the background stars \citep[see e.g.][]{Ch11,SL12,CL13,LC13}.
This could happen only during the last stages of inspiral. In our case the IMBHs do not attain such a configuration, thus the massive-perturbers' fast relaxation dominates the enhanced rate of TDEs by the MBH.

Besides their contribution to the TDE rate by the MBH, IMBHs can, by themselves, directly disrupt stars. The rate of TDEs by the IMBHs can be readily estimated by the rate of random close encounters of stars with the IMBHs.
\begin{equation}\label{eq:TDEI}
\Gamma_{I}=4\pi\int r^2 n_*(r)\sigma(r)v_*(r)n_I(r)dr,
\end{equation}
where $n_*(r)$ is the stellar number density, $\sigma(r)$ is the cross section for the IMBHs including gravitational focusing, evaluated using the tidal disruption radius of the IMBHs,
$v_*(r)$ is the stellar velocity as a function of the radius and $n_I(r)$ is the IMBHs number density.
For our system, at the end of the last merging, Equation (\ref{eq:TDEI}) gives $\Gamma_I\sim 10^{-3}$~yr$^{-1}$, a total rate of TDEs by all IMBHs, which is comparable with the one due to the MBH alone.
Thus, in this scenario we expect both TDEs by the MBH and by the IMBHs. Interestingly these two sources will produce different signatures \citep[see][and references therein]{Lo12}. The IMBHs
contribution can be long lived, since they apparently survive in the central {2}~pc and they do not inspiral to the MBH, nor are they ejected from nucleus. 

\subsection{Strong Mass Segregation and the Cusp Structure and Evolution}
\label{subs:sms}

As discussed in \S \ref{sec:res}, the central density cusp is already
established after the sixth infall. The density grows with the number
of infalls while the slope of the cusp remains almost
constant. Following the final merger of the ICs, the system evolves
due to two-body relaxation dominated by the IMBHs. The stellar
component slightly changes its slope, while the IMBHs distribution
maintains its initial slope.

Since the number density of IMBHs is much lower than that of stars we
expect the system to be in the strong mass segregation regime
\citep{AH09}. Indeed, the two populations settle to different density
profiles with different slopes. After rescaling the simulation time
with the relaxation time of the system, we find that it extends to
$\sim12\:{\rm Gyr}$.  At the end of the simulation, the system
distribution of the IMBHs (high-mass particles) is well fitted by a
power-law,
\begin{equation}
\rho_{H}(r)=6.20\times10^{3}r^{-2.32}\text{M}_{\odot}{\rm
pc}^{-3},\label{eq:cusp1}
\end{equation}
(see Figure \ref{fig:denIMBHs}). The slope of the IMBH cusp is, therefore,
within the range expected in the strong mass segregation case for the
higher mass particles ($2\lesssim\alpha_{H}\lesssim11/4$; see
\citealt{AH09}). 
{ Note that the IMBH density profile,
shown both in the right panel of Figures \ref{fig:den_ev} and \ref{fig:denIMBHs}, includes data on a small number of objects, and may therefore be sensitive to significant statistical noise, and the exact power-law profile calculated should be taken with caution. Nevertheless, all three simulations we have run show consistent results, and very similar power-law profiles clearly indicating a cusp profile for the IMBHs compared with the core profile of the stellar population.} 
The stellar component can be fitted by a shallower
radial profile,
\begin{equation}
\rho_{L}(r)=1.4\times10^{5}r^{-1.26},\label{eq:cusp2}
\end{equation}
whose power-law index is somewhat shallower than expected
($3/2\lesssim\alpha_{L}\lesssim7/4$).

These results qualitatively differ from those obtained for IMBH-free
ICs case for which we found the stellar population of the NSC to have an
external power-law distribution with a slope of $\sim-1.8$ outside an
inner core that shrinks with time but still exists out to $0.5$~pc at
the end of the simulation. With IMBHs, the NSC develops the central
power-law cusp early in its evolution.

\subsection{Mass Segregation, Compact Objects and Gravitational Wave Sources }

\label{sec:GWs} The inspiral of compact objects onto MBHs is a potential
source of gravitational waves (GWs). For MBHs with masses comparable
to Sgr A$^{*}$, the GW sources associated with extreme mass ratio
inspirals (EMRIs) might be detectable by future GW observatories, such
as the proposed Laser Interferometer Space Antenna
(eLISA\footnote{http:www.elisa.org/}, \citealt{Am07, Am11}). Stars are
tidally destroyed before they enter the Schwarzschild radius of the
MBH, while compact objects such as stellar black holes, neutron stars
and white dwarfs can survive to eventually produce EMRI GW events. IMBHs
may also be affected by GW emission during a close approach to MBH,
and their inspiral would have a much larger mass ratio, i.e. producing
``intermediate mass ratio inspiral'' (IMRI) events. Such inspirals are
among the most interesting GW sources that are
expected to be detected by eLISA.

The decay time due to GWs emission from an initial semimajor axis $a$
is given by
\begin{equation}
t_{GW}=\frac{5}{256}\frac{c^{5}}{G^{3}}\frac{a^{4}}{\mu M_{12}^{2}},
\end{equation}
where $\mu$ is the reduced mass of the system, while $M_{12}$ is the
mass of the binary \citep{Pe64,Pe08}. Figure \ref{fig:ecc} shows that
the IMBHs can reach very high eccentricities, and their evaluated GW
inspiral time can be much shorter than a Hubble time, but typically
much longer than the time between strong scattering by other IMBHs.  
Such
scatterings can increase the IMBH separation from the MBH. Our
simulations do not account for the GW emission, and therefore it is
not clear whether any of the IMBHs is likely to produce IMRI GW
sources. To check this effect we evaluated the ratio between the
analytically estimated change in energy due to GW emission, and the
energy variation due to scattering. This ratio, during the relaxation
of the system, is always $\ll1$, i.e. relaxation dominates the IMBH
dynamical evolution, and the IMBHs are unlikely to inspiral and become
GW sources. Future simulations with GW emission effects could shed
more light on this possibility.

As already found by \citet{Po04}, the stars most bound to an infalling
IMBH (which
are typically the  most massive, due to mass segregation in the IC),
can still accompany the IMBH as it inspirals to the NSC center, even
after the disruption of the cluster bulk. Each particle in our ICs is
$200\,{\rm M}_{\odot}$; we find that the total mass in bound stars
(typically 10 stars) brought in with each IMBH is
$\sim2000$~M$_{\odot}$.  Later on, as the IMBHs reach the central
region even these stars are typically stripped and left behind. IMBHs
can, therefore, induce a more extreme mass segregation than suggested by
the IMBH-free scenario \citep{PM14,An14}, and help refill the central
NSC core, and assist the development of the central stellar cusp. Since
IMBH-hosting ICs are expected to be mass segregated \citep{Ba77},
stellar black holes are likely to reside closest to the IMBH. These
stellar black holes can, therefore, be brought by the IMBHs to the NSC
center. Such compact objects in the closest regions to the MBH could
therefore potentially contribute to the EMRI GW rate \citep{HA06,OKL}.

\section{Summary and Conclusions}\label{sec:sum}

In this work we explored the formation and evolution of NSCs through
the merging of ICs containing IMBHs. The presence of IMBHs substantially
modifies the structure of the NSC. Our main results are as follows.

\begin{itemize}

\item Following the infall and disruption of an inspiraling cluster,
its IMBH continues to spiral in, accompanied by a small group of stars
or compact objects, which are later stripped from the IMBH close to
the central MBH. The IMBHs later scatter each other and the NSC stars,
and can develop high eccentricities.

\item In contrast to the IMBH-free case, the NSC develops a cusp
already in its early evolutionary stages.

\item The system is strongly mass segregated: IMBHs and stars evolve separately
and settle on two different quasi-stationary profiles characterized
by a central cusp with different slopes.
 
\item The NSC is tangentially anisotropic and oblate, very similar to simulated
NSCs without IMBHs.

\item The IMBHs decrease the relaxation time of the NSC by a factor of
a few hundred, leading to the fast refilling of the loss cone for tidal
disruption of stars by the central MBH. The system is in the full loss
cone regime and the tidal disruption rates are two orders of magnitude
larger than in an NSC without IMBHs. In addition the rate of TDEs
by the IMBHs themselves is comparable to that of the central MBH.

\item { The large rate of TDEs expected from IMBHs-hosting NSCs compared to the
much lower TDE rate estimates from observations, suggests that typical NSCs likely do not contain IMBHs. If cluster-infall is one of the main channels for NSC formation, this also implies
that the majority of massive clusters do not host IMBHs. We do caution, however, that the accuracy and reliability of current TDE rate estimates is still limited; and better estimates would be available from future surveys. }

\end{itemize}

\acknowledgments{ This research was supported by the I-CORE Program of
the Planning and Budgeting Committee and The Israel Science Foundation
grant 1829/12, and in part at the Technion by a fellowship from the
Lady Davis Foundation. The work was also supported by NSF grant
AST-1312034. }



\begin{thebibliography}

\bibitem[Agarwal \& Milosavljevic (2011)]{Ag11}Agarwal, M., \& Milosavljevic, M. 2011, ApJ, 729, 35
\bibitem[Alexander \& Hopman(2009)]{AH09}Alexander, T., \& Hopman, C. 2009, ApJ, 697, 1861
\bibitem[Amaro-Seoane et al.(2007)]{Am07}Amaro-Seoane P., Gair J. R., Freitag M., et al. 2007, CQGra, 24, R113
\bibitem[Amaro-Seoane \& Preto(2011)]{Am11}Amaro-Seoane, P., \& Preto, M. 2011, CQGra, 28, 094017
\bibitem[Antonini (2013)]{An13}Antonini, F. 2013, ApJ, 763, 62
\bibitem[Antonini (2014)]{An14}Antonini, F. 2014, ApJ, 794, 106
\bibitem[Antonini et al.(2012)]{An12}Antonini, F., Capuzzo-Dolcetta, R., Mastrobuono-Battisti, A., \& Merritt, D. 2012, ApJ, 750, 111 (Paper I)
\bibitem[Bahcall \& Wolf (1976)]{Ba76}Bahcall, J. N., \& Wolf, R. A. 1976, ApJ, 209, 214
\bibitem[Bahcall \& Wolf (1977)]{Ba77}Bahcall, J. N. \& Wolf, R. A. 1977, ApJ, 216, 883
\bibitem[B\"oker (2010)]{Bo10}B\"oker, T. 2010, in Astrophysics and Space Science Proc., The Impact of HST on European Astronomy, ed. W. Burton, L. L. Christensen, \& F. D. Macchetto (Berlin: Springer), 99
\bibitem[Bromm \& Loeb (2003)]{Br03}Bromm, V. \& Loeb, A. 2003, ApJ, 596, 34
\bibitem[Capuzzo-Dolcetta (1993)]{Ca93}Capuzzo-Dolcetta, R. 1993, ApJ, 415, 616
\bibitem[Capuzzo-Dolcetta \& Miocchi (2008a)]{Ca08a}Capuzzo-Dolcetta, R., \& Miocchi, P. 2008a, MNRAS, 388, L69
\bibitem[Capuzzo-Dolcetta \& Miocchi (2008b)]{Ca08b}Capuzzo-Dolcetta, R., \& Miocchi, P. 2008b, ApJ, 681, 1136
\bibitem[Casertano \& Hut~(1985)]{CH85} Casertano, S., Hut, P., 1985, ApJ, 298, 80
\bibitem[Chen et al.(2011)]{Ch11}Chen, X., Sesana, A., Madau, P., \& Liu, F. K. 2011, ApJL, 729, L13
\bibitem[Chen \& Liu(2013)]{CL13}Chen, X., \& Liu, F. K. 2013, ApJ, 762, 95
\bibitem[Ebisuzaki et al.(2001)]{Eb01}Ebisuzaki, T., Makino, J., Tsuru, T. G., et al. 2001, ApJL, 562, L19
\bibitem[Freitag et al.(2006)]{FGR06}Freitag, M., G\"urkan, M.~A. \& Rasio F.~A. 2006, MNRAS, 368, 141
\bibitem[Fujii et al.(2009)]{Fu09}Fujii, M., Iwasawa, M., Funato, Y., \& Makino, J. 2009, ApJ, 695, 1421
\bibitem[Gaburov et al.(2009)]{Ga09}Gaburov, E., Harfst, S., \& Portegies Zwart, S. 2009, NewA., 14, 630
\bibitem[Genzel et al.(2010)]{Ge10}Genzel, R., Eisenhauer, F., \& Gillessen, S. 2010, RvMP, 82, 3121
\bibitem[Gnedin et al.(2013)]{Gn13}Gnedin O. Y., Ostriker J. P., Tremaine S., 2014, ApJ, 785, 71
\bibitem[Graham \& Spitler(2009)]{GS09}Graham A.~W. \& Spitler L.~R. 2009, MNRAS, 397, 2148
\bibitem[Greene \& Ho (2008)]{Gr08}Greene, J. E., Ho, L. C., \& Barth, A. J. 2008, ApJ, 688, 159
\bibitem[Gualandris \& Merritt~(2009)]{Gu09} Gualandris, A., \& Merritt, D., 2009, ApJ, 705, 361
\bibitem[G\"urkan et al.(2004)]{GFR04}G\"urkan, M.~A., Freitag, M. \& Rasio, F.~A. 2004, ApJ, 604, 632 
\bibitem[Harfst et al.(2007)]{Ha07}Harfst, S., Gualandris, A., Merritt, D., et al. 2007, NewA, 12, 357
\bibitem[Hopman \& Alexander (2006)]{HA06}Hopman, C. \& Alexander, T. 2006, ApJL, 645, L133 
\bibitem[Jiang et al. (2011)]{Ji11}Jiang, Y., Greene, J. E., \& Ho, L. C. 2011, ApJL, 737, L45
\bibitem[Khabibullin \& Sazonov(2014)] {KS14} Khabibullin, I.\& Sazonov, S. 2014, MNRAS, 444, 1041
\bibitem[Komossa (2012)]{Ko12}Komossa, S. 2012, Tidal Disruption Events and AGN Outbursts, Madrid, Spain, Edited by R. Saxton; S. Komossa; EPJ Web of Conferences, 39, 02001
\bibitem[Launhardt et al.(2002)]{La02}Launhardt, R., Zylka, R., \& Mezger, P. G. 2002, A\&A, 384, 112
\bibitem[Liu \& Chen(2013)]{LC13}Liu, F. K., \& Chen, X. 2013, ApJ, 767, 18
\bibitem[Lodato(2012)]{Lo12}Lodato, G. 2012, Tidal Disruption Events and AGN Outbursts, Madrid, Spain, Edited by R. Saxton; S. Komossa; EPJ Web of Conferences, 39, 01001
\bibitem[Loeb \& Rasio (1994)]{Lo94}Loeb, A. \& Rasio, F. A. 1994, ApJ, 432, 52
\bibitem[Madau \& Rees (2001)]{Ma01}Madau, P. \& Rees, M. J. 2001, ApJL, 551, L27
\bibitem[Madigan \& Levin(2012)]{ML12}Madigan A.~M. \& Levin, Y. 2012, ApJ, 754, 42
\bibitem[Makino \& Taiji (1998)]{Ma98}Makino, J., \& Taiji, M. (ed.) 1998, Scientific Simulations with Special-purpose Computers. The GRAPE Systems (New York: Wiley-VCH), 248
\bibitem[Mastrobuono-Battisti \& Perets(2013)]{MBP13} Mastrobuono-Battisti, A., Perets, H.~B. 2013, ApJ, 779, 85 
\bibitem[McMillan \& Dehnen~(2005)]{MD05} McMillan, P. J., Dehnen, W., 2005, MNRAS, 363, 1205
\bibitem[Merritt (2010)]{Me10}Merritt, D. 2010, ApJ, 718, 739
\bibitem[Merritt (2013)]{Me13}Merritt, D. 2013, CQGra, 30, 24
\bibitem[Milosavljevi\'c (2004)]{Mi04}Milosavljevi\'c, M. 2004, ApJ, 605, L13
\bibitem[O'Leary, Kocsis \& Loeb(2009)]{OKL}O'Leary, R., Kocsis, B., \& Loeb, A.
2009, MNRAS, 395,2127
\bibitem[Peters(1964)]{Pe64}Peters P.~C. 1964, Phys. Rev., 136, 1224
\bibitem[Perets et al.(2007)]{Pe07}Perets, H.~B., Hopman, C., \& Alexander, T. 2007, ApJ, 656, 709
\bibitem[Perets \& Alexander(2008)]{Pe08}Perets, H. B., \& Alexander, T. 2008, ApJ, 677, 146
\bibitem[Perets \& Mastrobuono-Battisti(2014)]{PM14} Perets, H.~B. \& Mastrobuono-Battisti, A. 2014, ApJL, 784, L44
\bibitem[Portegies Zwart et al.(2004)]{Po04}Portegies Zwart, S. F., Baumgardt, H., Hut, P., Makino, J., \& McMillan, S. L. W. 2004, Nature, 428, 724
\bibitem[Portegies Zwart et al.(2006)]{Po06}Portegies Zwart, S. F., Baumgardt, H., McMillan, et al. 2006, ApJ, 641, 319
\bibitem[Scott \& Graham(2013)]{SG13} Scott, N. \& Graham, A.~W. 2013, ApJ, 763, 76 
\bibitem[Silk \& Arons(1975)]{Si75}Silk, J. \& Arons, J. 1975, ApJ, 200, L131
\bibitem[Stone \& Loeb(2012)]{SL12}Stone, N., \& Loeb, A. 2012, MNRAS, 422, 1933
\bibitem[Szell et al.~(2005)]{SM05} Szell, A., Merritt, D., \& Kevrekidis, I. G. 2005, PhRvL, 95, 081102
\bibitem[Tremaine et al. (1975)]{Tr75}Tremaine, S. D., Ostriker, J. P., \& Spitzer, L. 1975, ApJ, 196, 407
\end{thebibliography}
\end{document}